\documentclass[aps,pre,reprint]{revtex4-2}

\usepackage{amsmath}
\usepackage{amssymb}
\usepackage[separate-uncertainty]{siunitx}

\usepackage{graphicx}
\usepackage{chemformula}

\graphicspath{{figures/}}

\renewcommand{\vec}[1]{\boldsymbol{#1}}

\newcommand{\abs}[1]{\left\vert #1 \right\vert}
\newcommand{\real}[1]{\Re\left\{#1\right\}}

\DeclareSIUnit{\Molar}{M}
\DeclareSIUnit{\uM}{\micro\Molar}
\DeclareSIUnit{\mM}{\milli\Molar}

\begin{document}

\title{Measuring Colloidomer Hydrodynamics with Holographic Video Microscopy}

\author{Jatin Abacousnac}
\author{Jasna Brujic}
\author{David G. Grier}

\affiliation{Department of Physics and Center for Soft Matter Research, New York University, New York, NY 10003, USA}

\date{June 2023}

\begin{abstract}
    In-line holographic video microscopy records a wealth of information
    about the microscopic structure and dynamics of colloidal materials.
    Powerful analytical techniques are available to retrieve that
    information when the colloidal particles are
    well-separated. Large assemblies of close-packed particles
    create holograms that are substantially more challenging to interpret.
    We demonstrate that Rayleigh-Sommerfeld back-propagation
    is useful for analyzing holograms
    of colloidomer chains, close-packed linear assemblies of
    micrometer-scale emulsion droplets.
    Colloidomers are fully flexible chains and
    undergo three-dimensional configurational changes under the
    combined influence of random thermal forces and hydrodynamic
    forces. We demonstrate the ability of holographic reconstruction
    to track these changes as colloidomers sediment
    through water in a horizontal slit pore.
    Comparing holographically measured configurational trajectories
    with predictions of hydrodynamic models both validates
    the analytical technique for this valuable class of self-organizing
    materials and also provides insights into the influence of geometric
    confinement on colloidomer hydrodynamics.
\end{abstract}

\maketitle

\section{Introduction}
\label{sec:introduction}

\begin{figure*}
    \centering
    \includegraphics[width=0.9\textwidth]{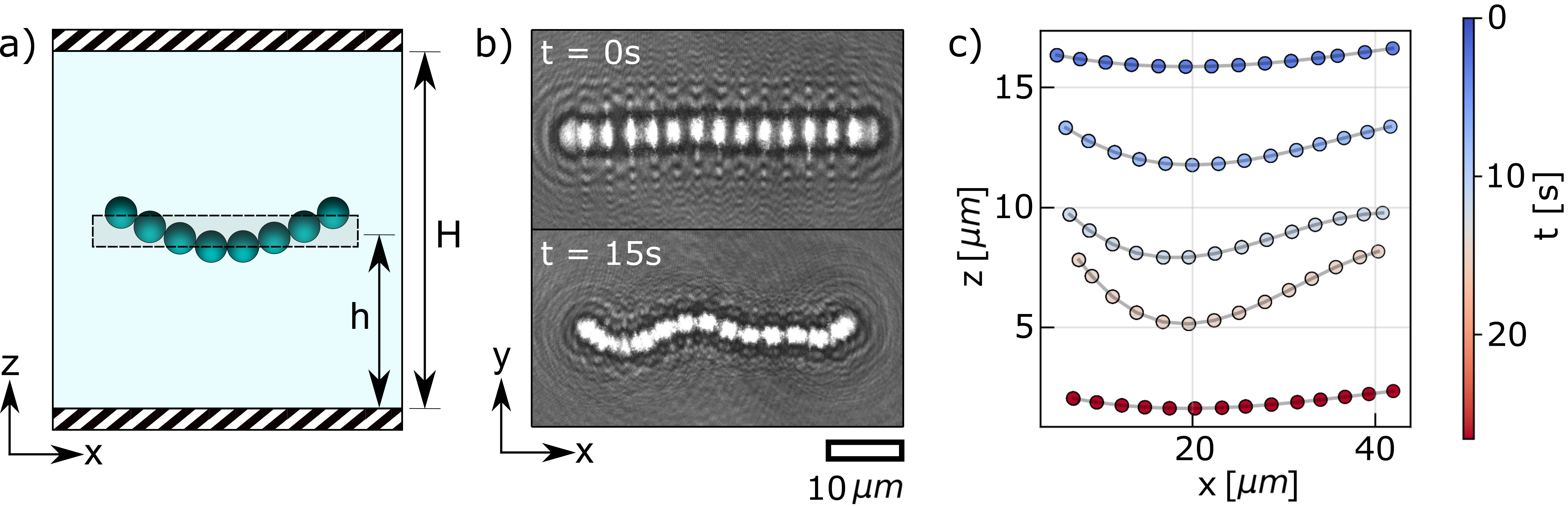}
    \caption{(a) Schematic view of a colloidomer's confined sedimentation. A chain of
    droplets is suspended in a layer of water between two horizontal glass surfaces
    separated by height $H$.
    The colloidomer is lifted along the vertical $\hat{z}$ axis
    and stretched to its full extent along $\hat{x}$
    by a pair of holographic optical tweezers.
    The traps then are extinguished, allowing the colloidomer to sediment
    freely.
    The chain bends and coils as it settles, and therefore is increasingly
    distorted relative to the ideal rod-like structure, shown at height $h$.
    (b) Holographic image of a model 14-droplet chain at two different times
    during its descent. (c) Trajectory of the chain from (b) obtained by analyzing
    holograms with Rayleigh-Sommerfeld backpropagation.
    Each point represents the three-dimensional position of one droplet, projected into the $x$-$z$ plane, and is colored by time.
    Hydrodynamic coupling to the lower wall suppresses
    curvature at late times.}
\label{fig:tryptich_schematic}
\end{figure*}

The conformational dynamics of a confined polymer is strongly
influenced by interactions with neighbors and
hydrodynamic coupling to bounding surfaces.
Such systems typically are studied through numerical simulation
because the relevant length and time scales are not
experimentally accessible.
This situation recently has changed with the introduction of
micrometer-scale colloidal analogs for molecular polymers.
These include chains of colloidal spheres
\citep{Byrom2014, Cunha2022} and emulsion droplets \citep{Feng2013, McMullen2018, McMullen2022},
linked by complementary DNA or nanoparticle
bridges.
These model soft-matter systems, known as colloidomers,
behave in much the same way as conventional
polymers with the advantage that they can be observed directly through optical
microscopy.
Experimental studies of colloidomer dynamics
not only offer insights into the
microscopic mechanisms of polymer physics, but also
clarify how colloidomers can be induced to self-organize
into hierarchical
structures with practical applications \cite{Coulais2018}.

The challenge in all such studies is to track the
individual beads in a colloidal chain as they move
relative to each other in three dimensions.
Conventional microscopy cannot readily follow objects
as they move out of the focal plane.
Confocal microscopy solves this problem, but is too slow to track conformational changes.
Lorenz-Mie analysis of holographic microscopy data
addresses both
of these issues \cite{lee2007characterizing},
but is impractical for complex assemblies of particles
\cite{fung2012imaging}, especially if their configuration changes over time.
Here, we introduce a general method based on holographic video
microscopy to measure the three-dimensional
conformational trajectory
of a colloidomer chain.
We demonstrate this technique with an
experimental study of fully flexible colloidomer chains
sedimenting through a viscous fluid between
two parallel hard walls.
In addition to validating the
technique, these measurements reveal the influence of
hydrodynamic coupling on the
dynamics of a flexible chain in quantitative
agreement with hydrodynamic models.

Our system is shown schematically in Fig.~\ref{fig:tryptich_schematic}(a).
Colloidomer chains are assembled from
monodisperse emulsion droplets composed of silicone oil
stabilized with lipid surfactant.
The population of droplets has a mean radius of
$a_p = \SI{1.58(4)}{\um}$, as determined
by Total Holographic Characterization
(xSight, Spheryx) \cite{lee2007characterizing}.
Droplets are linked into colloidomers using gold
nanoparticles to act as bridges and
are dispersed in a layer of water between two horizontal glass surfaces
that are separated by $H \approx \SI{24}{\um}$.
Links between droplets can flow freely across the droplets' surfaces.
Colloidomers therefore behave like freely jointed
chains \cite{McMullen2018}.
The sealed sample is mounted on the stage
of an in-line holographic video microscope that is
outfitted with holographic optical traps \cite{Lee2007,OBrien2019}.
Details of the colloidomers' synthesis and
characterization are provided in Appendix~\ref{sec:synthesis}.

Holograms of colloidomers, such as the examples
in Fig.~\ref{fig:tryptich_schematic}(b),
are analyzed with Rayleigh-Sommerfeld
back-propagation \cite{Dixon2011} to obtain the three-dimensional
position of each droplet in the chain.
These coordinates are then linked into the three-dimensional
configuration of the chain at each recorded
time step in its sedimentation, as shown in
Fig.~\ref{fig:tryptich_schematic}(c).
Measurements on chains of different lengths
reveal trends in sedimentation rate and
curvature that can be compared with
predictions based on models for the
forces acting on the chain.

Previous studies \cite{Li2013,Cunha2022,shashank2023dynamics} have noted that
hydrodynamic coupling among the beads in a
sedimenting chain reduces the viscous drag on
beads near the center relative
to those at the ends.
As a consequence, the central beads tend to sediment faster, causing
an initially linear chain to
bend into a hairpin.
The representative data in Fig.~\ref{fig:tryptich_schematic}(c) illustrate how confinement by rigid walls
suppresses the development of curvature, both at the beginning of the trajectory
and also at the end.
Measuring the dynamics of sedimenting
colloidomers allows us to quantify the chains' coupling
to the walls and thus to test idealized models
for bead-chain hydrodynamics.

\section{Tracking sedimenting chains with Rayleigh-Sommerfeld Backpropagation}
\label{sec:rayleighsommerfeld}

\begin{figure}
    \centering
    \includegraphics[width=\columnwidth]{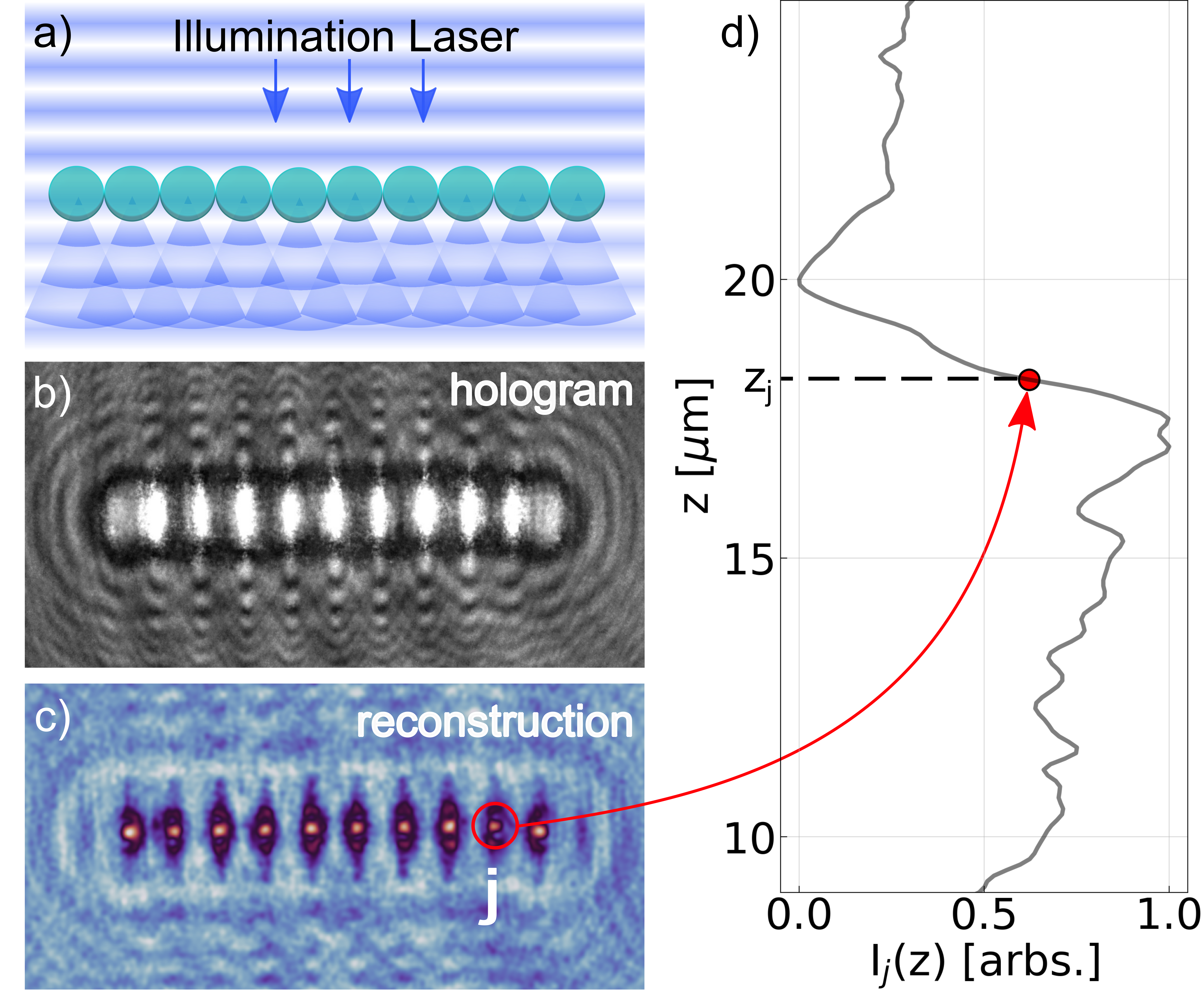}
    \caption{Holographic recording and reconstruction of a colloidomer's three-dimensional configuration.
    (a) A chain of droplets is illuminated with a collimated laser beam. Interference between
    the scattered and incident light creates an intensity pattern that is magnified by a microscope and recorded with a video camera.
    (b) Hologram of a 10-droplet colloidomer stretched to its full extent with holographic optical tweezers.
    (c) Rayleigh-Sommerfeld reconstruction of the scattered intensity, $I_R(\vec{r}, z)$, in the plane $z = \SI{18}{\um}$ above the microscope's focal plane.
    (d) Axial dependence of the reconstructed intensity, $I_j(z) = I_R(\vec{r}_j, z)$, at the in-plane position, $\vec{r}_j$,
    of one of the ten droplets. The inflection point is identified with the
    droplet's axial position, $z_j(t)$, at time $t$.
    Similar reconstructions are performed for each droplet in the chain.}
    \label{fig:reconstruction}
\end{figure}

Figure~\ref{fig:reconstruction} illustrates the measurement technique.
The three-dimensional configuration of a colloidomer is
recorded through in-line holographic video microscopy
\cite{sheng2006digital,Lee2007}.
As shown schematically in Fig.~\ref{fig:reconstruction}(a),
the sample is
illuminated with a collimated laser beam.
Light scattered by the droplets
interferes with the rest of the beam in the
focal plane of a microscope.
The intensity of the magnified interference
pattern is then recorded with a standard
video camera, as shown in Fig.~\ref{fig:reconstruction}(b).
Each such video image is a hologram of the
colloidomer and encodes information about
each droplet's size, refractive index and three-dimensional position \cite{Lee2007}.
Standard methods \cite{lee2007characterizing} to extract this information are confounded by the strongly overlapping
scattering patterns from neighboring droplets.
Rather than attempting to model such complex light-scattering patterns \cite{martin2022inline},
we instead localize the individual droplets
by numerically refocusing the hologram.

Given an estimate for the field scattered
by a colloidomer into the focal plane,
$E_s(\vec{r}, 0)$, we
can numerically reconstruct the
scattered field at
height $z$ above the focal plane with the Rayleigh-Sommerfeld
diffraction integral
\cite{Goodman2005,Lee2007,cheong2010strategies}:
\begin{subequations}
    \label{eq:rayleighsommerfeld}
\begin{equation}
    E_s(\vec{r}, z)
    =
    \int
    \tilde{E}_s(\vec{q}, 0) \, H(\vec{q}, -z)
    \, e^{i\vec{q}\cdot\vec{r}}
    \, d^2q,
\end{equation}
where
\begin{equation}
    \tilde{E}_s(\vec{q}, 0)
    =
    \int E_s(\vec{r}, 0) \,
    e^{-i \vec{q} \cdot \vec{r}} \,
    d^2 r ,
\end{equation}
and where
\begin{equation}
    \label{eq:rs_propagator}
    H(\vec{q},-z)
    = e^{iz(k^2 - q^2)^{\frac{1}{2}}}
\end{equation}
is the Fourier transform of the Rayleigh-Sommerfeld propagator
\cite{Goodman2005}.
The associated estimate for the scattered light's intensity at height $z$ is
\begin{equation}
    I_s(\vec{r}, z) = \abs{E_s(\vec{r}, z)}^2.
\end{equation}
\end{subequations}

The intensity distribution recorded by the camera
can be modeled as the superposition
of the electric field due to the incident plane wave, $\vec{E}_0(\vec{r})$,
and the scattered field in the
focal plane, $\vec{E}_s(\vec{r}, 0)$
\cite{Lee2007,Dixon2011}:
\begin{subequations}
\begin{align}
\label{eq:interference}
I(\vec{r}, 0)
& =
\abs{\vec{E}_0 (\vec{r}) + \vec{E}_s (\vec{r}, 0)}^2 \\
& \approx
I_0(\vec{r})
+ 2 \real{\vec{E}_0^\ast (\vec{r}) \cdot \vec{E}_s (\vec{r}, 0)},
\label{eq:model}
\end{align}
\end{subequations}
where $I_0(\vec{r}) = \abs{\vec{E}_0(\vec{r})}^2$
is the intensity of the illumination in the
absence of scatterers and where
we assume that
the intensity of the scattered wave is
small enough compared with the other terms to be
omitted.
If, furthermore, the scatterer is far enough from the focal plane
that polarization rotations can be neglected,
the normalized hologram,
\begin{subequations}
    \label{eq:b}
\begin{equation}
    b(\vec{r})
    \equiv
    \frac{I(\vec{r}, 0)}{I_0(\vec{r})} - 1,
\end{equation}
is related to the scattered field
in the focal plane by
\begin{equation}
    b(\vec{r})
    \approx
    \frac{E_s(\vec{r}, 0)}{E_0^\ast(\vec{r})} +
    \frac{E_s^\ast(\vec{r}, 0)}{E_0(\vec{r})}.
\end{equation}
\end{subequations}
This expression can be further simplified
by modeling the collimated illumination
as a plane wave aligned with the imaging plane
so that $E_0^\ast(\vec{r}) = E_0(\vec{r}) = E_0$
\cite{moyses2013robustness}.

Equation~\eqref{eq:b} shows that
the normalized hologram
records a superposition of the
field scattered by the
colloidomer and also a fictive
field propagating in the opposite
direction, which corresponds to
the colloidomer's mirror image in the focal plane.
The associated twin image can be suppressed numerically
\cite{latychevskaia2007solution,zhang2018twin,Dixon2011}
at the cost of additional computational complexity.
If the colloidomer is far enough from the focal plane, however,
perturbations due to the twin image
are weak enough to ignore without further processing
\cite{Dixon2011}.
In that case, the scattered intensity, $I_R(\vec{r}, z)$,
at height $z$ above the focal plane can
be reconstructed numerically from
a measured hologram
by substituting $b(\vec{r})$ for
$E_s(\vec{r}, 0)$ in Eq.~\eqref{eq:rayleighsommerfeld}.

The image in Fig.~\ref{fig:reconstruction}(c) shows
an estimate for the scattered intensity,
$I_R(\vec{r}, z)$, obtained from the hologram in Fig.~\ref{fig:reconstruction}(b)
in the plane $z = \SI{18}{\um}$.
Because this colloidomer has been stretched into a straight line
by optical tweezers, all of the droplets come into sharpest
focus in the same plane and appear in the reconstructed image
as an array of small bright spots.
Interestingly, these intensity maxima are offset from
the brightest regions in the original hologram, which
illustrates the diffractive nature of the image-formation
process.

A quantitative estimate for a droplet's
three-dimensional position
starts with an estimate for its
in-plane position, $\vec{r}_p$, which might
be obtained manually.
The axial position, $z_p$,
is then estimated by seeking the point of inflection
in the numerically reconstructed
axial intensity profile \cite{Lee2007}
along $\vec{r}_p$,
as shown in Fig.~\ref{fig:reconstruction}(d).
The image is then refocused to $z_p$ and the
estimate for $\vec{r}_p$ is refined
using standard particle-tracking
algorithms \cite{Crocker1996}.
Finally, $z_p$ is refined using the
improved estimate for $\vec{r}_p$.
Numerical uncertainties in the fit values
suggest that this protocol achieves
a precision
of $\Delta r_p = \SI{0.1}{pixel} = \SI{5}{\nm}$ for the droplet's in-plane position
and $\Delta z_p = \SI{1}{pixel} = \SI{48}{\nm}$ along the axial
direction.

The same procedure can be performed for each droplet
in a chain and yields a snapshot of the
colloidomer's instantaneous configuration.
The mean separation between droplet
centers,
\begin{equation}
    \Delta r
    =
    \frac{1}{N-1}
    \sum_{j = 1}^{N-1}
    \abs{\vec{r}_j(t) - \vec{r}_{j+1}(t)},
\end{equation}
should be equal to the
diameter of a single droplet.
The data in Fig.~\ref{fig:tryptich_schematic}
yield
$\Delta r/2 = \SI{1.5(1)}{\um}$,
which is consistent
with the holographically
measured radius,
$a_p = \SI{1.58(4)}{\um}$, for
the population of emulsion droplets
used to assemble the chain.
This agreement helps to validate
the tracking technique.

The data in Fig.~\ref{fig:tryptich_schematic}(c)
show the projection of a
colloidomer's three-dimensional configuration
onto the $x$-$z$ plane
as it freely sediments within its sample cell.
Such configurational trajectories can be
compared with predictions of hydrodynamic
models both to validate the tracking technique
and also to cast new light on the influence
of confinement on colloidomer dynamics.

\section{Sedimentation of confined chains of beads}
\label{sec:hydrodynamics}

\subsection{Sedimentation of a rigid chain}
\label{sec:rigid}

The drag force experienced by an isolated
sphere of radius $a_p$ moving
with velocity $\vec{v}$ through a  viscous fluid
is given by \cite{HappelBrenner1991},
\begin{subequations}
\begin{gather}
    \vec{F}(\vec{v})
    = - \gamma_0 \, \vec{v}, \quad \text{where} \\
    \label{eq:stokes}
    \gamma_0 = 6 \pi \eta \, a_p
\end{gather}
\end{subequations}
is the Stokes drag coefficient
in a fluid of viscosity $\eta$.
A fully extended chain of spheres sedimenting
perpendicularly to its axis can be modeled as
a long ellipsoid,
for which the drag coefficient is
\cite{HappelBrenner1991}
\begin{equation}
\label{eq:ellipse}
    \gamma_\infty
    =
    \frac{8\pi\eta \, a}{
    \ln\left(2\frac{a}{b}\right) + \frac{1}{2}},
\end{equation}
where $a$ and $b$ are the semi-major and semi-minor axes,
respectively.
Combining Eqs.~\eqref{eq:stokes} and \eqref{eq:ellipse}
suggests that the
drag coefficient for an extended chain of $N$ spheres
is approximately
\begin{equation}
\label{eq:rod_drag_nowall}
    \gamma_N
    =
    \frac{4}{3}
    \left(
    \frac{N}{\ln{2N}+\frac{1}{2}}
    \right)
    \gamma_0
\end{equation}
in the limit $N \gg 1$.
The denominator of Eq.~\eqref{eq:rod_drag_nowall}
implicitly incorporates hydrodynamic coupling among the
spheres, albeit without allowing the spheres to
rearrange themselves.

\subsection{Confinement by a slit pore}
\label{sec:confinement}

Hydrodynamic coupling to bounding surfaces
enhances the drag on a moving object.
For a sphere moving perpendicularly to a rigid wall,
Fax\'en shows that \cite{Brenner1962,HappelBrenner1991}
\begin{equation}
\label{eq:faxen}
    \gamma(h)
    \approx
    \gamma_0 \left(1 - \frac{9}{8} \frac{a_p}{h} \right)^{-1},
\end{equation}
where $h$ is the distance from the center of the sphere to the wall.
The analogous result for a horizontal chain of spheres
at height
$h$ above a horizontal wall is \cite{Hadamard1911}
\begin{equation}
\label{eq:rod_drag_onewall}
    \gamma_N(h)
    =
    \gamma_N
    \left(
    1 -
    \frac{9}{8} \frac{a_p}{h} \frac{\gamma_N}{\gamma_0}\right)^{-1}.
\end{equation}
Adding a second parallel wall at distance $H$ from the first, as
shown in Fig.~\ref{fig:tryptich_schematic}(a),
disproportionately increases the complexity of the drag calculation \cite{Liron1976}.
If $H \gg a_p$, however, we may invoke Oseen's linear superposition
approximation \cite{Dufresne2000},
\begin{equation}
    \gamma_N(h, H)
    \approx
    \gamma_N(h) + \gamma_N(H-h) - \gamma_N.
\end{equation}
With these simplifying approximations,
a horizontally extended colloidomer
sedimenting through a viscous fluid
in the space between two horizontal plane walls
should fall with a speed that depends on
its height in the channel,
\begin{subequations}
\label{eq:prediction}
\begin{gather}
    v_p(h)
    =
    \frac{N}{\gamma_N(h, H)} \, \Delta m_p g,
    \quad \text{where}\\
    \Delta m_p
    =
    \frac{4}{3} \pi \, a_p^3 \,
    \Delta \rho_p
\end{gather}
\end{subequations}
is the buoyant mass of a single droplet
given its buoyant mass density,
$\Delta \rho_p$,
and where $g = \SI{9.81}{\meter\per\square\second}$
is the acceleration due to gravity.

\section{Results}
\label{sec:results}

\subsection{Measured sedimentation of confined colloidomers}

\begin{figure}
    \centering
    \includegraphics[width=\columnwidth]{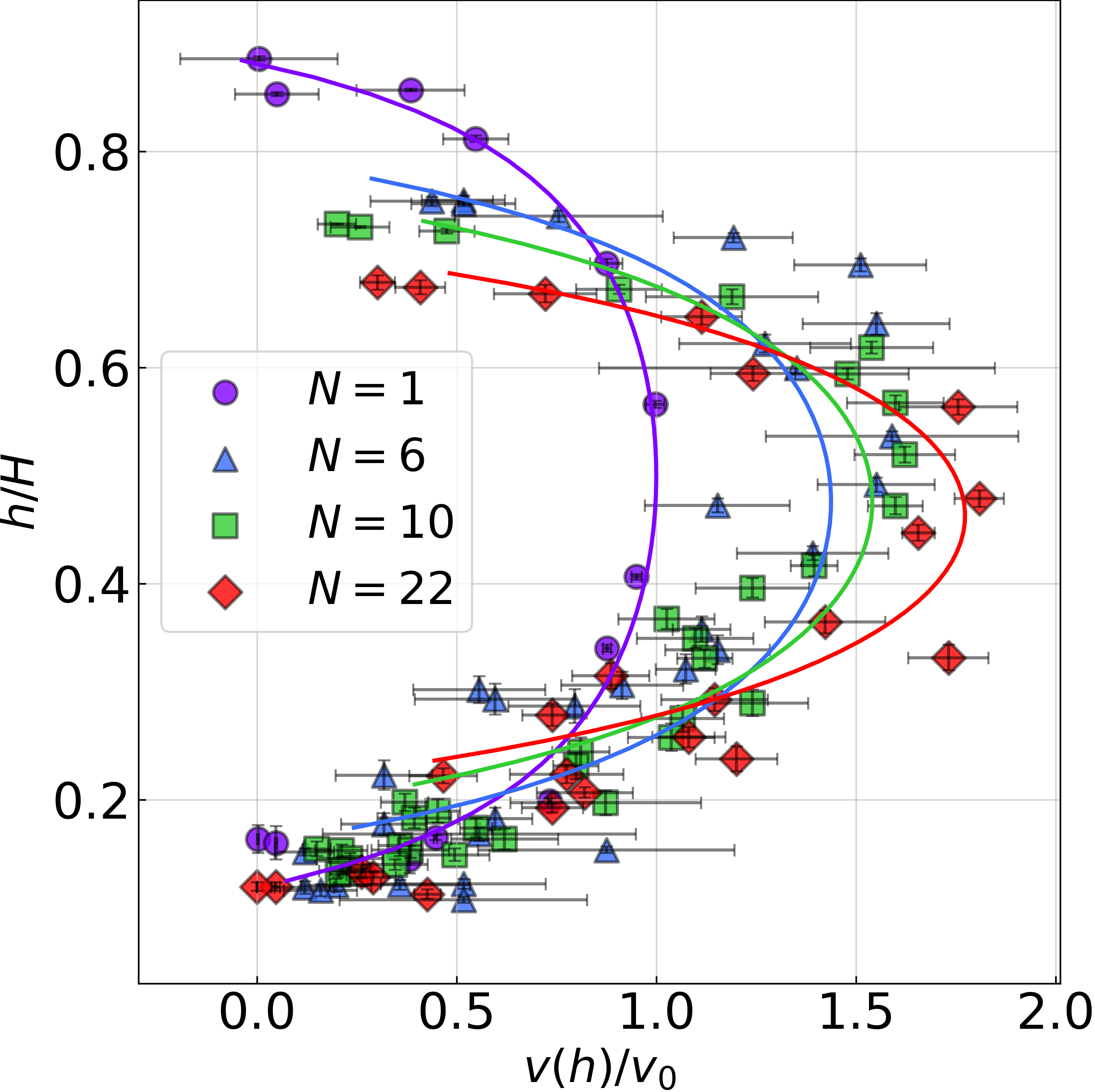}
    \caption{Velocity profiles of sedimenting colloidomers of chain lengths $N = 6$,
    10 and 22.
    Monomer data ($N = 1$) are fit to Eq.~\eqref{eq:prediction} with
    $\gamma_N = \gamma_0$ . The fit yields the droplet's
    buoyant density,
    $\Delta\rho_p = \SI{166(7)}{\kg\per\cubic\meter}$, and the wall
    separation, $H$. Colloidomer data are
    fit to Eq.~\eqref{eq:prediction} using the same value
    for the droplet density, leaving
    only $H$ as an adjustable parameter.}
    \label{fig:velocity_profile}
\end{figure}

Figure~\ref{fig:velocity_profile} presents
measurements of colloidomers' sedimentation
speed, $v(h)$, as a function of height
in the channel,
$h$, and the number of droplets in the chain, $N$.
A measurement is initiated by trapping the
two ends of a colloidomer with a pair of holographic optical tweezers, stretching the chain to its maximum extension, lifting it to a height $h$ above the sample chamber's lower wall,
and then extinguishing the traps to let the chain
sediment freely.
The optical traps are powered by a fiber laser
(IPG Photonics, YLR-LP-SF) operating at a vacuum
wavelength of \SI{1064}{\nm}.
The light is formed into programmable optical traps
by imprinting it with computer-generated holograms
using a liquid-crystal spatial light modulator (Holoeye PLUTO)
\cite{Reicherter1999,curtis2002dynamic,Polin2005,OBrien2019}.
The holograms are projected through the microscope's objective
lens using the standard holographic optical trapping technique \citep{Grier2003,Dufresne1998} and are controlled using
the pyfab interface
\cite{grier2024pyfab}.

In-line holographic imaging is performed with
a \SI{3}{\milli\watt} collimated beam
provided by a fiber-coupled diode laser (Coherent Cube) operating at a vacuum wavelength of $\lambda = \SI{447}{\nm}$.
The beam's \SI{3}{\mm} waist barely overfills the input pupil of the
microscope's objective lens
(Nikon Plan Apo, 100\texttimes,
numerical aperture 1.4, oil immersion).
A small proportion of the incident light
is scattered by the chain of droplets
and interferes with the remainder of the beam
in the focal plane of the objective lens.
The magnified interference pattern is relayed
to a monochrome video camera (FLIR Flea3 USB 3.0) that
records its intensity every $\Delta t = \SI{66}{\ms}$.
The exposure time of \SI{100}{\us} is fast enough
to avoid distortions due to motion blurring
\cite{Dixon2011_02}.

Each holographic snapshot in the video stream is analyzed with the methods of Sec.~\ref{sec:rayleighsommerfeld}.
A colloidomer's height, $h(t)$, at time
$t$ is computed as the average of the
axial positions of its constituent droplets.
The uncertainty in the height consists of
the standard deviation of the droplets'
axial positions
in quadrature with
the numerical error in the single-droplet
measurements.
The colloidomer's sedimentation speed is
computed as the change in height during
the camera's frame interval, $\Delta t$.
All speeds are scaled by the
sedimentation speed for a single
sphere at the midplane,
\begin{align}
    v_0
    =
    \frac{\Delta m_p g}{\gamma_0}
    \left( 1 - \frac{9}{2} \frac{a_p}{H} \right),
\end{align}
which is found to be
 $v_0 = \SI{0.79(0.02)}{\um\per\second}$
by fitting to the unscaled data for $N = 1$.
Figure~\ref{fig:velocity_profile} presents
the scaled results for the single droplet.

The single-droplet fit yields the buoyant density of
the silicone oil mixture,
$\Delta \rho_p = \SI{166(7)}{\kg\per\cubic\meter}$,
and the height of the
channel, $H$.
The former is consistent with independent
measurements using a
density meter (DMA 4500 M, Anton Paar)
and is treated as a fixed parameter
when analyzing the data for colloidomers.
Because different sample cells have slightly
different heights, $H \approx \SI{25}{\um}$ is
treated as an adjustable parameter
for each experiment and is resolved to
within \SI{100}{\nm}.

The solid curves in Fig~\ref{fig:velocity_profile}
are single-parameter fits
to Eq.~\eqref{eq:prediction}
for chains of length $N = 6$, 10 and 22.
Predictions of the rigid-chain
model agree well with the
measured sedimentation profiles of
colloidomers despite the
colloidomers' flexibility.

Colloidomers sediment more slowly near
walls than
individual spheres because of the
length-dependent
enhancement to
hydrodynamic coupling described
by Eq.~\eqref{eq:rod_drag_onewall}.
This effect is stronger for longer chains.
Conversely, colloidomers sediment more rapidly than
individual spheres near the middle of the channel
where wall-induced drag has less influence than
the drag reduction afforded by coupling
to neighboring droplets.
The crossover between these two effects is
modeled by the length-dependent
denominator in Eq.~\eqref{eq:rod_drag_nowall}.
None of these comparisons, however, take
account of the curvature that develops
as colloidomers sediment.

\subsection{Curvature of confined sedimenting
colloidomers}
\label{sec:curvature}

Hydrodynamic coupling reduces the drag
on droplets near the center of the chain
relative to those at the ends.
The central droplets therefore should
sediment faster, causing the chain to bend,
and eventually to adopt a hairpin configuration
\citep{Li2013,Cunha2022}.
The tendency for the center of a flexible chain
to sediment fastest is evident in the early stages of the trajectory plotted in Fig.~\ref{fig:tryptich_schematic}(c).

Hydrodynamic
coupling to the upper wall complicates
this process because the
associated contribution to the drag
tends to be greater for spheres near the
center of the chain, which have more
neighbors.
The additional drag tends to suppress the emergence
of curvature relative to a free chain.
Once curvature develops, however, this
contribution to the wall-associated
drag will most strongly affect the
droplets that remain closest to the wall,
accelerating the evolution of curvature.
Depending on chain length and its initial
distance from the upper wall, this competition
could favor the development of
negative curvature,
with the
end droplets initially falling faster than
the central droplets \cite{perrin2019peeling}.

\begin{figure}
    \centering
    \includegraphics[width=\linewidth]{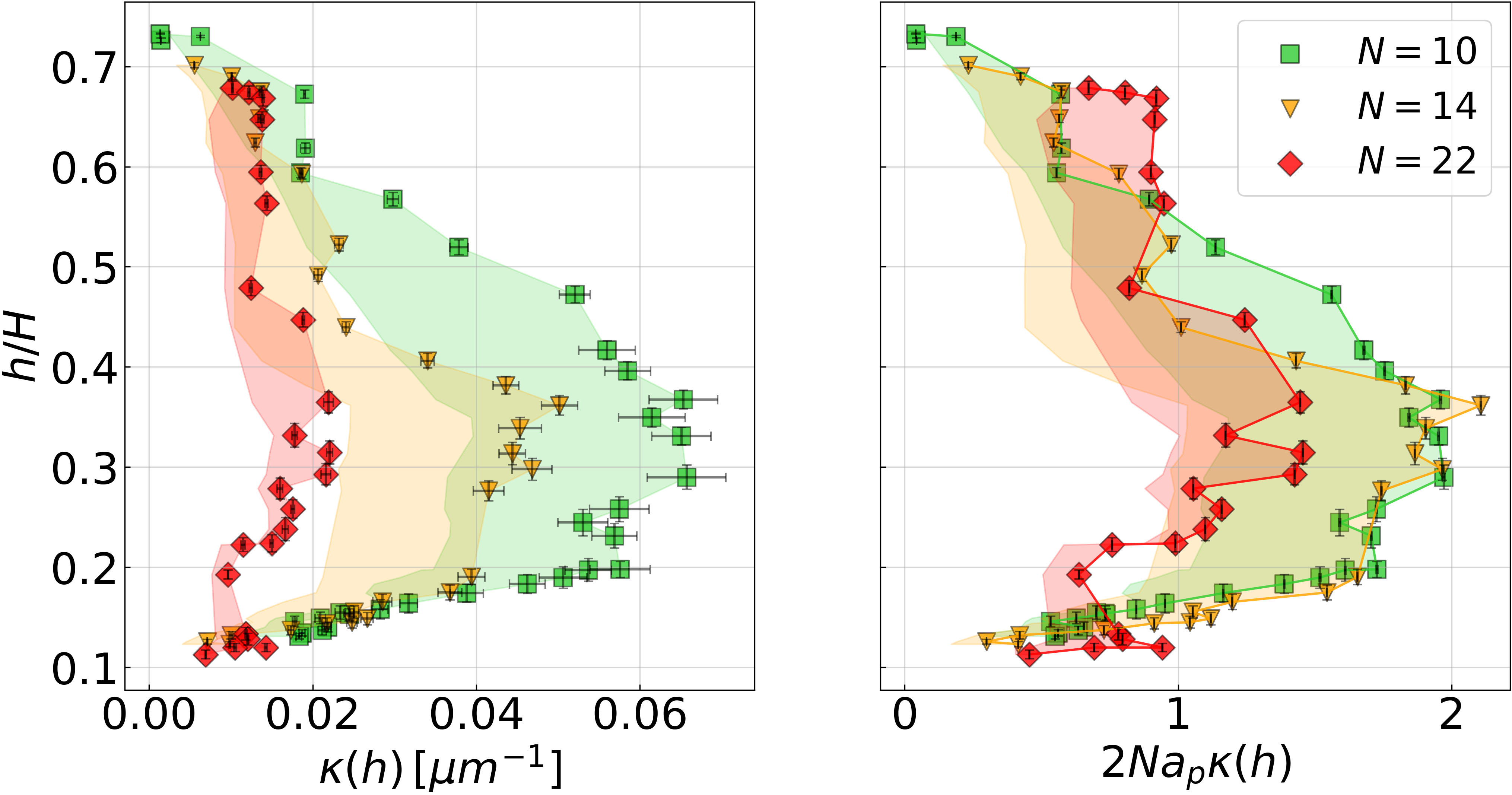}
    \caption{(a) The maximum three-point curvature of a colloidomer chain, $\kappa(h)$, as a function of the colloidomer's mean height in the channel. Data are shown for $N = 10$ (green squares), $N=14$ (gold inverted triangles), and $N=22$ (red diamonds). Shaded areas show the range between mean and maximum values.
    (b) Scaling the curvature by the chain length  effectively collapses the three
    data sets.}
    \label{fig:curvature_profile}
\end{figure}

We avoid this complication
by releasing the chains away
from the upper wall thereby weakening that wall's
influence on the chains' dynamics and focusing attention
on the lower wall's influence.
As indicated in Fig.~\ref{fig:tryptich_schematic}(c),
colloidomers are stretched to their
full extent with optical tweezers, lifted to height
$h = \num{0.7} H$ in the channel and then
released to sediment freely.
Figure~\ref{fig:curvature_profile}
shows how a colloidomer's curvature
evolves as it sediments.
The instantaneous curvature at arc length
$s$ along the time-evolving three-dimensional conformation, $\vec{r}(s,t)$,
may be computed as
\begin{equation}
\label{eq:curvature}
    \kappa(s,t)
    =
    \frac{\abs{\vec{r}' \times \vec{r}''}}{
    \abs{\vec{r}'}^3},
\end{equation}
where primes denote derivatives with respect to
$s$.
Equation~\eqref{eq:curvature}
can be approximated for the discrete
vertices of a colloidomer by the three-point
Menger curvature.
The discrete points in Fig.~\ref{fig:curvature_profile}(a)
report the maximum curvature,
$\kappa(h) = \max\{\kappa(s, h)\}$,
for colloidomers of length $N = 10$, 14 and 22,
where $h = h(t)$ is the
colloidomer's mean height in the channel
at time $t$.
The shaded regions indicate the range of curvature values
between the mean curvature and the maximum.

Introducing curvature into a colloidomer involves
moving droplets longitudinally along the chain
to conserve the chain's length.
The associated contribution to the drag is proportional to the
total amount of longitudinal displacement
and therefore is proportional to the length of
the chain.
Scaling the curvature by the chain length
therefore yields a dimensionless curvature
that should increase linearly with time as
the chain sediments \cite{Cunha2022}.
The data collapse in Fig.~\ref{fig:curvature_profile}(b)
is consistent with this interpretation under conditions
where the chains are near the channel's midplane
and confinement effects are weak.

Hydrodynamically mediated curvature is augmented
by Brownian coiling \cite{McMullen2018},
whose influence can be seen in Fig.~\ref{fig:tryptich_schematic}(b).
The curvature due to coiling can exceed that for
hydrodynamic bending in a long chain that sediments slowly.
The curvature plotted in Fig.~\ref{fig:curvature_profile}
therefore becomes increasingly noisy
at later times, especially for the longest chain with $N = 22$.

As the chain sediments below the midplane
of the channel, its hydrodynamic coupling
to the lower wall increases.
This tends to reduce its hydrodynamically
induced curvature even before the
bottom-most droplets touch down.
Once the chain makes contact, the vertical
component of its curvature is mechanically
suppressed, leaving only the in-plane
curvature due to coiling.

\section{Discussion}

Holographic microscopy is a
powerful tool for measuring the three-dimensional trajectories of micrometer-scale objects \cite{Lee2007,lee2007characterizing}.
Taking a targeted approach to Rayleigh-Sommerfeld reconstruction extends these
capabilities to systems, such as colloidomers,
that consist of close-packed particles.
Measurements of colloidomer sedimentation in
a slit pore validate the technique by yielding
results for chains' trajectories that
are consistent with hydrodynamic models.

\begin{acknowledgments}
    This work was supported by the National
    Science Foundation through Award Number DMR-2104837 and Award Number DMR-2105255.
    The xSight holographic
    particle characterization instrument was acquired as shared
    instrumentation with support from the MRSEC program of the NSF under Award No.~DMR-1420073. The integrated holographic trapping and holographic microscopy instrument used for this study was constructed as shared instrumentation with support from the NSF under Award No. DMR-0922680.
    We gratefully acknowledge extensive
    discussions with Angus McMullen regarding
    the preparation of colloidomer chains.
\end{acknowledgments}
\appendix

\section{Colloidomer Synthesis}
\label{sec:synthesis}

Our emulsion droplets are created from a 1:1:1 mixture by volume fraction of three types of silane monomers: (a) dimethyldiethoxysilane (Sigma-Aldrich), commonly known as DMDES, (b) (3,3,3-trifluoropropyl) methyldimethoxysilane (Gelest), and (c) 3-glycidoxypropylmethyldiethoxysilane (Gelest).
Droplets are condensed out of an aqueous monomer solution
by ammonia-catalyzed hydrolysis and condensation following
the procedure described in Ref.~\cite{Elbers2015}.
Specifically, \SI{175}{\uL} of monomer is dissolved
in \SI{3.115}{\milli\liter} of deionized water by
vortexing. Droplet nucleation is initiated by adding
\SI{1}{\percent v/v} of \SI{27}{\percent} ammonium hydroxide
solution (Sigma-Aldrich).
Droplet growth is stopped after 6 hours
by washing the sample
with \SI{1}{\mM} sodium dodecyl sulfate solution
(SDS, Sigma-Aldrich).
The sample is centrifuged and resuspended with \SI{1}{\mM} SDS solution three times.
The completed emulsion is stored in \SI{5}{\mM} SDS.

The droplets are stabilized with lipid surfactants.
Lipid labeling is achieved by diluting \SI{10}{\uL} of packed droplets with \SI{190}{\uL} of \SI{5}{\mM} SDS solution containing \SI{1}{\uL} DSPE-PEG-SH thiol-terminated lipid (Avanti Polar Lipids, MW 2000). The mixture is kept on a rotary mixer overnight. Excess lipid is washed off by centrifuging, and replacing the supernatant layer with \SI{1}{\mM} SDS each time.

Droplets are assembled into linear chains using methods
described in Refs.~\cite{Obey1994,Elbers2015,McMullen2021}.
The emulsion droplets are first
concentrated by centrifugation before being redispersed in a commercial ferrofluid
(EMG 707, Ferrotec)
along with \SI{1}{\mM} Tris-EDTA (Thermo Fischer Scientific) at pH 8
and \SI{30}{\nm}-diameter gold nanoparticles (BBI Solutions, EMGC30) at a concentration
of \SI{e11}{\per\mL}.
This mixture is imbibed into a rectangular capillary tube (VitroCom) by capillary action
and is subjected to an external magnetic
field of roughly \SI{1}{\tesla} aligned
with the capillary's long axis
that is applied with a pair of
neodymium rare-earth magnets.
The emulsion droplets act as diamagnetic bubbles in the ferrofluid and are organized
into chains aligned with the field by
induced-dipole interactions.

Individual droplets become decorated with
gold nanoparticles that bind irreversibly
to the lipids' terminal thiol groups.
Once bound, the gold nanoparticles diffuse
across the droplets' fluid surfaces until
they reach the contact region between
neighboring droplets.
There, they form bridges
between the droplets, binding the
field-induced lines of droplets
into permanent colloidomer chains.
The droplets remain in a fluid state
so that the inter-droplet bonds
are fully flexible.
Hematite nanoparticles from the ferrofluid
are prevented from binding to the thiol groups
through the addition of EDTA.

Stable colloidomer chains form after roughly two hours at room temperature, after which they
are washed out of the capillary
into \SI{100}{\uL} of \SI{0.5}{\percent v/v} Pluronic F-108
solution (Sigma-Aldrich).
Excess ferrofluid is pipetted off the top with the help of a bar magnet. The remaining sample is pipetted between a glass microscope slide and a glass \#1.5 coverslip (Globe Scientific), and sealed using UV-cured adhesive (LOON, UV Clear Fly Finish).

%

\end{document}